\documentclass[11pt]{article}

\usepackage[paper=a4paper,dvips,top=2.0cm,left=2.2cm,right=2.2cm,bottom=2.9cm,footskip=1.4cm]{geometry}
\usepackage{amsfonts,amsmath,amsthm,amssymb}
\numberwithin{equation}{section}  
\usepackage[svgnames]{xcolor}
\usepackage{fix-cm}
\usepackage{sectsty}
\usepackage{fancyhdr}
\usepackage{lastpage}
\usepackage{graphicx}
\usepackage{url}
\usepackage{enumerate}
\usepackage{paralist}
\usepackage{multicol}
\usepackage{color}  
\usepackage{float}  
\usepackage{epsfig}
\usepackage{hyperref}   
\hypersetup{colorlinks=true,linkcolor=beamer@PRD, citecolor=beamer@PRD}
\usepackage{authblk}  
\usepackage{cite}  

\newcommand\myref[1]{\textcolor{beamer@PRD}{(}\ref{#1}\textcolor{beamer@PRD}{)}}

\definecolor{beamer@blue}{RGB}{0,0,255}
\definecolor{beamer@mediumblue}{RGB}{0,0,190}
\definecolor{beamer@midnightblue}{RGB}{25,25,112}
\definecolor{beamer@navy}{RGB}{0,0,128}
\definecolor{beamer@darkblue}{RGB}{0,0,139}
\definecolor{beamer@purple}{RGB}{128,0,128}
\definecolor{beamer@levander}{RGB}{100.,149.,237.}
\definecolor{beamer@PRD}{RGB}{46,48,146}
\definecolor{beamer@green}{RGB}{0,128,0}
\definecolor{beamer@darkgreen}{RGB}{0,100,0}
\definecolor{beamer@olive}{RGB}{128,128,0}
\definecolor{beamer@darkolivegreen}{RGB}{85,107,47}
\definecolor{beamer@gray}{RGB}{190,190,190}
\definecolor{beamer@ivry}{RGB}{220,220,220}
\definecolor{beamer@new}{RGB}{40,120,50}
\definecolor{shadecolor}{RGB}{220,220,220}
\definecolor{beamer@darkslategray}{RGB}{47,79,79}
\definecolor{beamer@chocolate}{RGB}{210,105,30}
\definecolor{beamer@brown}{RGB}{165,42,42}
\definecolor{beamer@orangered}{RGB}{255,69,0}
\definecolor{beamer@maroon}{RGB}{128,0,0}
\definecolor{beamer@white}{RGB}{234,242,243}
\definecolor{beamer@silver}{RGB}{0.5,0.45,0.37}

\begin{document}


\title{\textbf{Noncommutative $q$-photon added coherent states}}


\author{\textbf{Sanjib Dey$^{1,2}$ and V\'eronique Hussin$^{1,3}$}}
\date{\footnotesize{$^{1}$Centre de Recherches Math{\'e}matiques, Universit{\'e} de Montr{\'e}al, Montr{\'e}al--H3C 3J7, Qu{\'e}bec, Canada \\ $^{2}$Department of Mathematics and Statistics, Concordia University, Montr{\'e}al--H3G 1M8, Qu{\'e}bec, Canada \\ $^{3}$Department de Math{\'e}matiques et de Statistique, Universit{\'e} de Montr{\'e}al, Montr{\'e}al--H3C 3J7, Qu{\'e}bec, Canada \\ \small{E-mail: dey@crm.umontreal.ca, veronique.hussin@umontreal.ca}}}
\maketitle
    	
\thispagestyle{fancy}
\begin{abstract}
We construct the photon added coherent states of a noncommutative harmonic oscillator associated to a $q$-deformed oscillator algebra. Various nonclassical properties of the corresponding system are explored, first, by studying two different types of higher order quadrature squeezing, namely the Hillery-type and the Hong--Mandel-type and, second, by testing the sub-Poissonian nature of photon statistics in higher order with the help of the correlation function and the Mandel parameter. By comparing our results with those of the usual harmonic oscillator, we notice that the quadratures and photon number distributions in noncommutative case are more squeezed for the same values of the parameters and, thus, the photon added coherent states of noncommutative harmonic oscillator may be more nonclassical in comparison to the ordinary harmonic oscillator.
\end{abstract}	 
\begin{section}{Introduction} \label{sec1}
\addtolength{\footskip}{-0.1cm} 
\addtolength{\voffset}{1.2cm} 
It is familiar that a coherent state $\vert\alpha\rangle$ is an exceptional type of quantum state that resembles a classical light field very closely and exhibits Poisson photon number distribution with an average photon number $\vert\alpha\vert^2$. Coherent states have well-defined amplitude and phase, with minimal fluctuations permitted by the Heisenberg uncertainty principle. On contrary, a Fock state $\vert n\rangle$ is entirely quantum mechanical, and in the language of Glauber and Sudarshan \cite{Glauber2,Sudarshan2}, it possesses purely \textit{nonclassical} features. 

The \textit{photon added coherent states} (PACS) are the intermediate states between the coherent state and the Fock state \cite{Agarwal_Tara}, which are obtained by the successive action ($m$ times) of the canonical creation operator $a^\dagger$ on the standard coherent state $\vert\alpha\rangle$, as given by
\begin{equation}\label{PhotnAdded}
\vert\alpha,m\rangle =\frac{1}{\mathcal{N}(\alpha,m)}a^{\dagger m}\vert\alpha\rangle,
\end{equation}
with the normalisation constant being $\mathcal{N}^2(\alpha,m) = \langle\alpha\vert a^ma^{\dagger m}\vert\alpha\rangle$. The PACS \myref{PhotnAdded} have been studied by many authors in different contexts \cite{Dodonov_etal,Sivakumar_PACS,Sixdeniers_Penson,Daoud,Popov,Naderi_etal,Sudheesh_etal,Sudheesh_etal1, Duc_Noh,Gorska_Penson_Duchamp,Safaeian_Tavassoly,Mojaveri_Dehghani_Mahmoodi,Mojaveri_Dehghani_Mohammadzadeh}. In \cite{Sivakumar_PACS}, it was shown that the states \myref{PhotnAdded} can be realised as eigenstates of the annihilation operator, $f(\hat{n},m)a\vert\alpha,m\rangle=\alpha\vert\alpha,m\rangle$, with $f(\hat{n},m)=1-m/(1+\hat{n})$, such that $\vert\alpha,m\rangle$ can be considered as nonlinear coherent states. The overcompleteness of PACS was explicitly shown in \cite{Sixdeniers_Penson} by using the Stieltjes power moment problem. The dynamical squeezing was investigated in \cite{Dodonov_etal}. Properties of nonlinear PACS have been explored in \cite{Safaeian_Tavassoly,Mojaveri_Dehghani_Mohammadzadeh} in different contexts. PACS have various applications in quantum optics, quantum information and computation, not only because they are nonclassical, but also because they can generate the entangled states \cite{Li_Jing_Zhan,Berrada_etal}. Like many other nonclassical states; such as, squeezed states \cite{Stoler,Hollenhorst,Walls}, Schr\"odinger cat states \cite{Dodonov_Malkin_Manko,Xia_Guo}, pair coherent states \cite{Agarwal_Biswas} and binomial states \cite{Stoler_Saleh_Teich,Lee}, the special features of PACS consist of quadrature squeezing, sub-Poissonian photon statistics, negativity in Wigner function, etc. PACS are produced in the interaction of a two-level
atom, with a cavity field initially prepared in the coherent state \cite{Agarwal_Tara}. Proper experiments behind the existence of a single-PACS \cite{Zavatta_Viciani_Bellini,Barbieri_etal} and two-PACS \cite{Kalamidas_Gerry_Benmoussa} have been performed successfully.  

The goal of the present manuscript is to introduce the formalism for the construction of PACS for a noncommutative harmonic oscillator (NCHO) originating from a $q$-deformed algebra, and to explore various nonclassical properties of the corresponding system. A detailed analysis of different types of squeezing properties in higher orders shows an improved degree of nonclassicality of the given system than that of an ordinary harmonic oscillator. This motivates us to explore the possibility that the noncommutative systems might be implemented for the purpose of quantum information processing with additional degrees of freedom, especially when we notice that the given NCHO might be realised physically. 

Our manuscript is organised as follows: In section \ref{sec2}, we show how to build a NCHO out of a $q$-deformed oscillator algebra and, then, we construct the PACS of the corresponding system. In section \ref{sec3}, we analyse the properties of two types of higher order quadrature squeezing, namely, the Hillery-type and the Hong--Mandel-type and provide a comparative analysis of our results between the NCHO and the ordinary harmonic oscillator. In section \ref{sec4}, we study the sub-Poissonian nature of photon statistics of our system by analysing the behaviour of the correlation function and the Mandel parameter in higher order.  Our conclusions are stated in section \ref{sec5}. 
\end{section}
\begin{section}{$q$-deformed PACS for NCHO}\label{sec2}
Following Refs. \cite{Arik_Coon,Biedenharn,Macfarlane,Sun_Fu,Kulish_Damaskinsky}, we start by considering a one-dimensional $q$-deformed oscillator algebra for the deformed annihilation and creation operators $A_q$ and $A_q^\dagger$ in the form
\begin{equation}\label{qDeformed}
A_qA_q^\dagger -q^2A_q^\dagger A_q=1, \qquad \text{for}~\vert q\vert < 1.
\end{equation}
The Fock space of the corresponding algebra \myref{qDeformed} can be defined by choosing $q$-deformed integers $[n]_q$ in such a way that the following relations hold
\begin{eqnarray}\label{qFock}
\vert n\rangle_q &:=& \frac{A_q^{\dagger n}}{\sqrt{[n]_q!}}\vert 0\rangle_q, \qquad [n]_q!:= \displaystyle\prod_{k=1}^{n}[k]_q, \qquad [0]_q! :=1, \\
~[n]_q &:=& \frac{1-q^{2n}}{1-q^2}, \qquad A_q\vert 0\rangle_q = 0, \qquad ~_{q}\langle 0 \vert 0\rangle_q = 1. \notag
\end{eqnarray}
It immediately follows that the operators $A_q$ and $A_q^\dagger$ act as lowering and raising operators, respectively, in the deformed space
\begin{eqnarray}\label{qLadder}
A_q\vert n \rangle_q &=& \sqrt{[n]_q}~\vert n-1\rangle_q, \\
A_q^\dagger\vert n \rangle_q &=& \sqrt{[n+1]_q}~\vert n+1\rangle_q. \notag
\end{eqnarray}
It means that the states $\vert n\rangle_q$ form an orthonormal basis in the $q$-deformed Hilbert space $\mathcal{H}_q$ spanned by the vectors $\vert\psi\rangle :=\sum_{n=0}^{\infty}c_n\vert n\rangle_q$ with $c_n\in\mathbb{C}$, such that $\langle\psi\vert\psi\rangle = \sum_{n=0}^{\infty}\vert c_n\vert^2<\infty$. Therefore, the commutation relation between $A_q$ and $A_q^\dagger$ is realised as follows
\begin{equation}
[A_q,A_q^\dagger]=1+(q^2-1)A_q^\dagger A_q=1+(q^2-1)[n]_q.
\end{equation}
The concept that the $q$-deformed algebras of type \myref{qDeformed} can be implemented for the construction of $q$-deformed harmonic oscillators was given by many authors \cite{Arik_Coon,Biedenharn,Macfarlane,Sun_Fu,Kulish_Damaskinsky}. Here we recall Refs. \cite{Bagchi_Fring,Dey_Fring_Gouba,Kempf_Mangano_Mann,Dey_Fring_Gouba_Castro} to construct a NCHO from the given algebra \myref{qDeformed}, instead. For this, we first express the deformed observables $X$ and $P$ in terms of the ladder operators $A_q, A_q^\dagger$ in the following form
\lhead{Noncommutative $q$-photon added coherent states}
\chead{}
\rhead{}
\begin{equation}\label{XP}
X=\gamma(A_q^\dagger+A_q) \qquad \text{and} \qquad P=i\delta(A_q^\dagger-A_q).
\end{equation}  
Thereafter, by choosing the appropriate constraints on the parameters, $\gamma =\sqrt{\hbar/(4m\omega)}, \delta = \sqrt{m\omega\hbar}$, and by using \myref{qDeformed} we obtain the following commutation relation between the position and momentum variables
\begin{equation}\label{NCOM}
[X,P]=\frac{2i\hbar}{1+q^2}+i\frac{q^2-1}{q^2+1}\left(2m\omega X^2+\frac{P^2}{2m\omega}\right).
\end{equation}
Notice that in the limit $q\rightarrow 1$, the commutator \myref{NCOM} reduces to the standard canonical commutation relation. The interesting feature of such type of noncommutative space-time \myref{NCOM} is that it leads to the existence of a minimal length as well as a minimal momentum \cite{Bagchi_Fring,Dey_Fring_Gouba}, which are the direct consequences of string theory. Furthermore, there exists a concrete self-adjoint representation of the ladder operators \cite{Dey_Fring_Gouba_Castro}
\begin{eqnarray}\label{HermitianRep}
A_q &=& \frac{i}{\sqrt{1-q^2}}\left(e^{-i\hat{x}}-e^{-i\hat{x}/2}e^{2\tau \hat{p}}\right), \\
A_q^\dagger &=& \frac{-i}{\sqrt{1-q^2}}\left(e^{i \hat{x}}-e^{2\tau \hat{p}}e^{i\hat{x}/2}\right), \notag
\end{eqnarray}
in terms of the canonical coordinates $x,p$ satisfying $[x,p]=i\hbar$, with $\hat{x}=x\sqrt{m\omega/\hbar}$ and $\hat{p}=p/\sqrt{\hbar m\omega}$  being dimensionless observables, and the deformation parameter $q$ being parametrised to $q=e^\tau$. It follows that the observables \myref{XP}, which satisfy \myref{NCOM} are Hermitian with respect to the representation \myref{HermitianRep}, i.e. $X^\dagger =X, P^\dagger = P$. As obviously, our representation \myref{HermitianRep} is not unique and with further investigations it may be possible to find other Hermitian representations. However, for our purpose it is important that there exists at least one such representation providing a self-consistent description of a physical system. In addition, we notice that our representation \myref{HermitianRep} is invariant under the simultaneous operation of parity $\mathcal{P}$ and time-reversal $\mathcal{T}$ operator, $\mathcal{PT}: x\rightarrow -x, p\rightarrow p, i\rightarrow -i$ \cite{Bender_Boettcher, Mostafazadeh_2002}. It means that the $\mathcal{PT}$-symmetry is inherited in the canonical variables on the noncommutative space \myref{XP}, $\mathcal{PT}: X\rightarrow -X, P\rightarrow P, i\rightarrow -i$, which indicates that any quantum model consisting of the observables $X, P$ may possess entirely real eigenvalues. Indeed, in some of the recent articles by one of the authors \cite{Dey_Fring_Gouba,Dey_Fring_Khantoul}, many similar type of noncommutative systems have been solved with real energy spectrum. It should be mentioned that in recent days various non-Hermitian but $\mathcal{PT}$-symmetric systems have been utilised for experiments in optical laboratory in many different directions, one may follow, for instance, \cite{El-Ganainy,Guo,Longhi,Chong_Ge_Cao_Stone}. 

Similar type of systems that we are discussing here have been used for the purpose of construction of different type of quantum optical models, for instance, coherent states \cite{Manko_Marmo_Sudarshan_Zaccaria,Quesne,Dey_Fring_squeezed,Dey_Fring_Gouba_Castro}, cat states \cite{Mancini,Osland,Dey,Dey_Fring_Hussin,Fakhri_Hashemi} and squeezed states \cite{Dey_Hussin}. Here we construct the PACS associated to the algebra \myref{qDeformed} by following the original definition \myref{PhotnAdded} as given by
\begin{eqnarray}\label{qPACS}
\vert\alpha,m\rangle_q &=& \frac{1}{\mathcal{N}(\alpha,m,q)}A_q^{\dagger m}\vert\alpha\rangle_q \notag \\
&=& \frac{1}{\mathcal{N}(\alpha,m,q)\mathcal{N}(\alpha,q)}\displaystyle\sum_{n=0}^{\infty}\frac{\alpha^n}{[n]_q!}\sqrt{[n+m]_q!}~\vert n+m\rangle_q, \quad \alpha\in\mathbb{C},
\end{eqnarray}
with the normalisation constant
\begin{equation}
\mathcal{N}^2(\alpha,m,q) ={}_q\langle\alpha,m\vert A_q^mA_q^{\dagger m}\vert\alpha,m\rangle_q=\frac{1}{\mathcal{N}^2(\alpha,q)}\displaystyle\sum_{n=0}^{\infty}\frac{\vert\alpha\vert^{2n}}{[n]_q!^2}[n+m]_q!,
\end{equation}
where $\vert\alpha\rangle_q=(1/\mathcal{N}(\alpha,q))\sum_{n=0}^{\infty}(\alpha^n/\sqrt{[n]_q!})\vert n\rangle_q$ is a standard $q$-deformed nonlinear coherent state \cite{Arik_Coon,Dey_Fring_Gouba_Castro,Dey}, with $\mathcal{N}^2(\alpha,q)=\sum_{n=0}^{\infty}\vert\alpha\vert^{2n}/[n]_q!$. Note that, the deformed PACS represented in \myref{qPACS} are absolutely general, which can be associated to any kind of $q$-deformed systems corresponding to the $q$-deformed integers $[n]_q$. In our case, we choose a particular form of $[n]_q$ as given in \myref{qFock} and utilise it in rest of our analysis. However, in the limit $q\rightarrow 1$, the entire structure reduces to the original PACS as proposed in \cite{Agarwal_Tara}. $q$-deformed photon added and photon depleted coherent states have been studied before in \cite{Naderi_etal}, however, in a completely different context. They utilised a different type of $q$-boson, called the inverse $q$-boson, which originates from the $q$-deformed algebra introduced in \cite{Quesne}.
\end{section} 
\begin{section}{Higher order squeezing in quadrature components}\label{sec3}
We now analyse various nonclassical properties of the deformed PACS \myref{qPACS} in the following two sections. In this section, we check whether the quadrature components for deformed PACS are squeezed or not. There exist two different types of higher order quadrature squeezing in the literature, Hillery-type \cite{Hillery} and Hong--Mandel-type \cite{Hong_Mandel}. These two types of higher order squeezing have been studied  for different quantum states in different contexts \cite{Gerry_Rodrigues,Zhang_Xu_Chai_Li,Zhan,Du_Gong,An,Sudheesh_etal1, Prakash_Mishra,Duc_Noh,Verma_Pathak,Aeineh_Tavassoly}. Let us discuss both of them in our case.  
\begin{subsection}{Hillery-type higher order squeezing}
According to Hillery \cite{Hillery}, a quadrature $Y_N(\phi)$ of the form
\begin{equation}
Y_N(\phi) = \frac{1}{2}(A_q^Ne^{-iN\phi}+A_q^{\dagger N}e^{iN\phi}),
\end{equation}
is said to be squeezed for an arbitrary order $N$ for the state $\vert\alpha,m\rangle_q$ for some values of $\phi$ if
\begin{equation}\label{QSqueezing}
_q\big\langle\alpha,m\big\vert (\Delta Y_N(\phi))^2\big\vert\alpha,m\big\rangle_q < \frac{1}{4} ~_q\big\langle\alpha,m\big\vert [A_q^N,A_q^{\dagger N}]\big\vert\alpha,m\big\rangle_q,
\end{equation}
where $\phi$ is the angle in the complex plane and $_q\langle (\Delta Y_N(\phi))^2\rangle_q={}_q\langle (Y_N(\phi))^2\rangle_q-{}_q\langle (Y_N(\phi))\rangle_q^2$ being the variance of the quadrature $Y_N(\phi)$. The LHS of \myref{QSqueezing} can be computed by utilising the following equations 
\begin{alignat}{1}
_q\langle (Y_N(\phi))^2\rangle_q &=\frac{1}{4}\left[{}_q\langle A_q^{2N}\rangle_qe^{-2iN\phi}+{}_q\langle A_q^{\dagger 2N}\rangle_qe^{2iN\phi}+{}_q\langle A_q^NA_q^{\dagger N}\rangle_q+{}_q\langle A_q^{\dagger N}A_q^N\rangle_q\right], \\
_q\langle (Y_N(\phi))\rangle_q^2 &=\frac{1}{4}\left[{}_q\langle A_q^{N}\rangle_q^2e^{-2iN\phi}+{}_q\langle A_q^{\dagger N}\rangle_q^2e^{2iN\phi}+2\left\vert {}_q\langle A_q^N\rangle_q\right\vert^2\right].
\end{alignat}
A degree of Hillery-type squeezing can be obtained by computing the squeezing coefficient $S_H$ as defined by
\begin{equation}\label{SH}
S_H =\frac{4~_q\langle (\Delta Y_N(\phi))^2\rangle_q-{}_q\langle [A_q^N,A_q^{\dagger N}]\rangle_q}{_q\langle [A_q^N,A_q^{\dagger N}]\rangle_q},
\end{equation}
where $S_H<0$ corresponds to the existence of squeezing. The squeezing is guaranteed if the relation \myref{QSqueezing} holds (and $S_H<0$) at least for the order $N=1$, however, we are interested to present a fairly general expression of the squeezing coefficient for an arbitrary order $N\geq 1$. For this, we compute
\begin{equation}\label{ADNAL}
_q\langle A_q^{\dagger N} A_q^L\rangle_q=\left\{ \begin{array}{lcl}
\frac{\alpha^{\ast (N-L)}}{{\mathcal{\hat{N}}^2}}\displaystyle\sum_{n=0}^{\infty}\frac{\vert\alpha\vert^{2n}[n+m]_q![n+m+N-L]_q!}{[n]_q![n+N-L]_q![n+m-L]_q!} & \mbox{if}
& N>L \\ \frac{\alpha^{L-N}}{{\mathcal{\hat{N}}^2}}\displaystyle\sum_{n=0}^{\infty}\frac{\vert\alpha\vert^{2n}[n+m]_q![n+m+L-N]_q!}{[n]_q![n+L-N]_q![n+m-N]_q!} & \mbox{if} & L>N,\end{array}\right.
\end{equation}
and
\begin{equation}\label{ANADL}
_q\langle A_q^N A_q^{\dagger L}\rangle_q=\left\{ \begin{array}{lcl}
\frac{\alpha^{N-L}}{{\mathcal{\hat{N}}^2}}\displaystyle\sum_{n=0}^{\infty}\frac{\vert\alpha\vert^{2n}[n+m+N]_q!}{[n]_q![n+N-L]_q!} & \mbox{if}
& N>L \\ \frac{\alpha^{\ast (L-N)}}{{\mathcal{\hat{N}}^2}}\displaystyle\sum_{n=0}^{\infty}\frac{\vert\alpha\vert^{2n}[n+m+L]_q!}{[n]_q![n+L-N]_q!} & \mbox{if} & L>N,\end{array}\right.
\end{equation}
with $\mathcal{\hat{N}}=\mathcal{N}(\alpha,m,q)\mathcal{N}(\alpha,q)$ and $N, L\in \mathbb{Z}$. Since, we will always use the normal ordered form of the operators, we do not require \myref{ANADL} in our analysis, however, for the sake of completeness of our calculations we present it explicitly. Note that the equations \myref{ADNAL} and \myref{ANADL} are valid for $\vert N-L\vert\geq 1$. For the case of $N=L$, we obtain
\begin{alignat}{1}
_q\langle A_q^{\dagger N}A_q^N\rangle_q &= \frac{1}{\mathcal{\hat{N}}^2}\displaystyle\sum_{n=0}^{\infty} \frac{\vert\alpha\vert^{2n}([n+m]_q!)^2}{([n]_q!)^2[n+m-N]_q!}, \label{AAD} \\
_q\langle A_q^N A_q^{\dagger N}\rangle_q &= \frac{1}{\mathcal{\hat{N}}^2}\displaystyle\sum_{n=0}^{\infty} \frac{\vert\alpha\vert^{2n}[n+m+N]_q!}{([n]_q!)^2}. \label{ADA}
\end{alignat}
In addition, we notice from \myref{ADNAL} and \myref{ANADL} that ${}_q\langle A_q^{\dagger N} A_q^L\rangle_q ={}_q\langle A_q^{\dagger L} A_q^N\rangle_q^\ast$ and ${}_q\langle A_q^N A_q^{\dagger L}\rangle_q ={}_q\langle A_q^L A_q^{\dagger N}\rangle_q^\ast$ and, therefore, the expectation values of $A_q^N$ and $A_q^{2N}$ are obtained by taking the complex conjugates of the mean values of $A_q^{\dagger N}$ and $A_q^{\dagger 2N}$, respectively. Thus, the $N$th order squeezing coefficient acquires the form
\begin{figure}
\centering   \includegraphics[width=8.2cm,height=6.0cm]{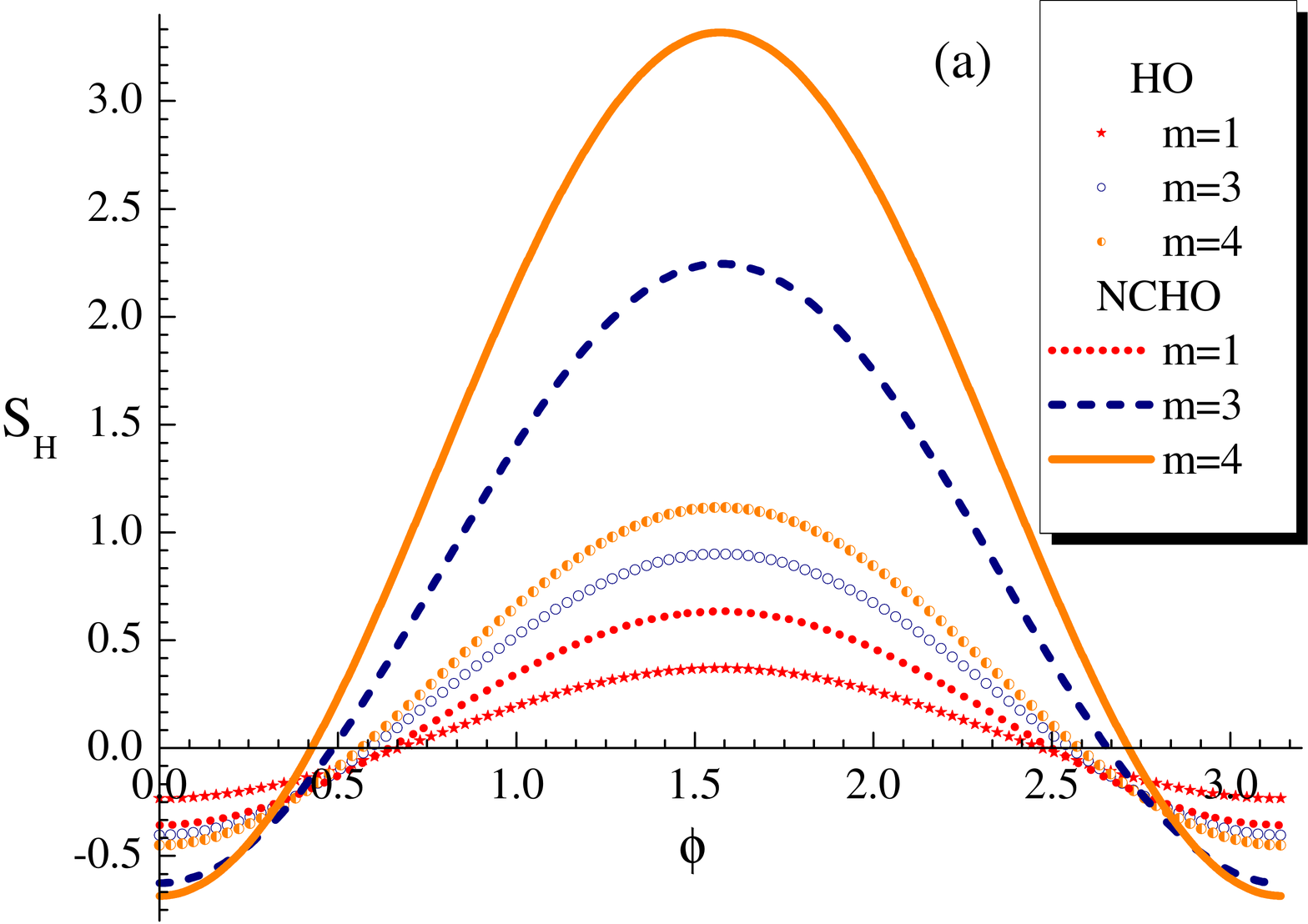}
\includegraphics[width=8.2cm,height=6.0cm]{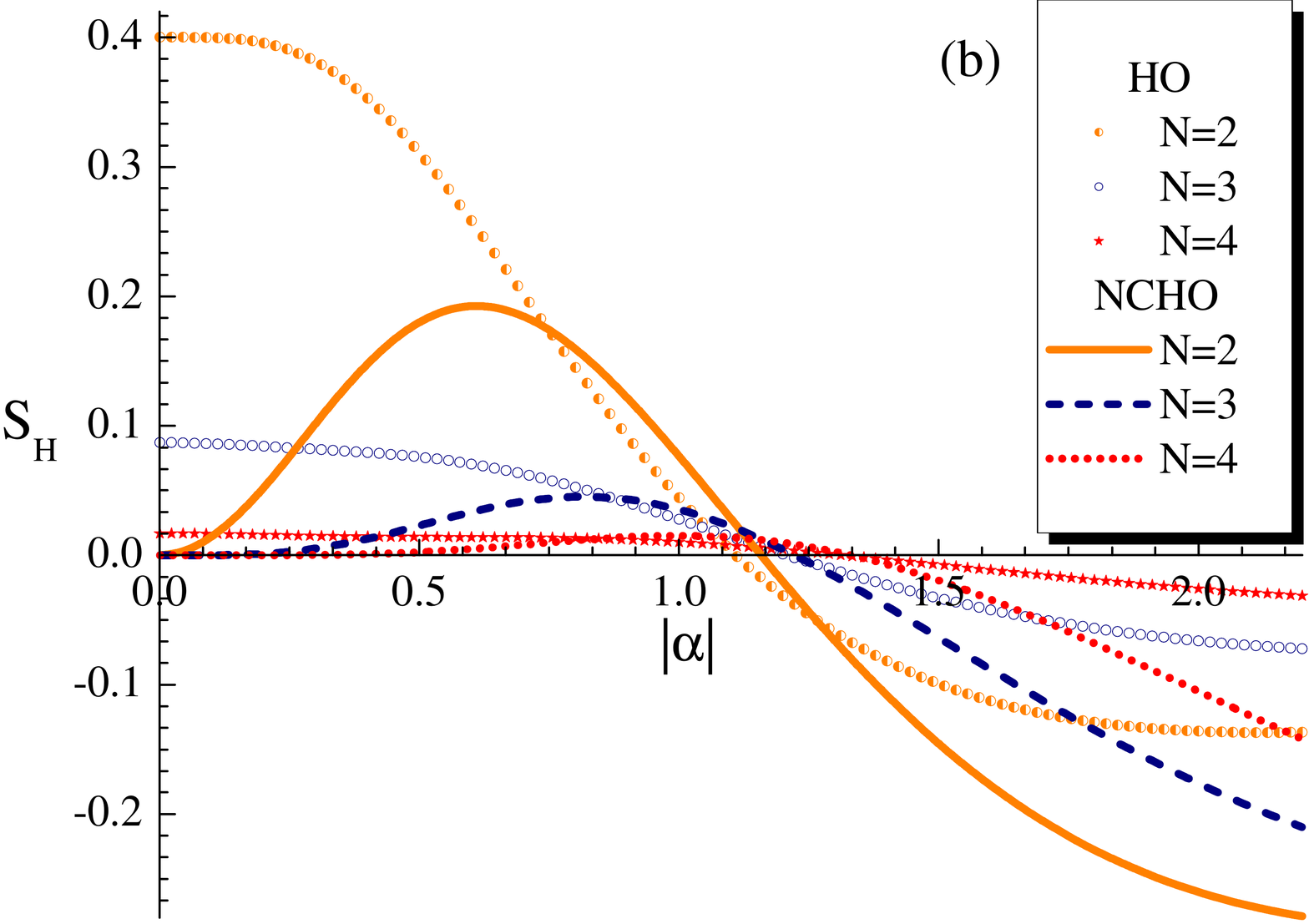}
\caption{\small{Comparison of Hillery-type higher order squeezing coefficient of a harmonic oscillator (scattered dots) versus a NCHO (for $q=0.9$) (solid and dotted lines) as a function of (a) $\phi$ for different values of $m$, with $N=1,\alpha =2.1$, (b) $\vert\alpha\vert$ for different values of $N$, with $\phi =0.1,m=1$.}}
\label{fig1}
\end{figure}
\begin{equation}\label{HillerySqueezing}
S_H = 2\frac{\text{Re}\left[\left({}_q\langle A_q^{2N}\rangle_q-{}_q\langle A_q^N\rangle_q^2\right)e^{-2iN\phi}\right]-\left\vert {}_q\langle A_q^N\rangle_q\right\vert^2+{}_q\langle A_q^{\dagger N}A_q^N\rangle_q}{_q\langle A_q^NA_q^{\dagger N}\rangle_q-{}_q\langle A_q^{\dagger N}A_q^N\rangle_q},
\end{equation}
which can be evaluated easily with the help of \myref{ADNAL}, \myref{AAD} and \myref{ADA}. In Fig. \ref{fig1}(a), we plot the first order $(N=1)$ squeezing coefficient \myref{HillerySqueezing} as a function of angle $\phi$ for different values of the added photon numbers $m$ and for a particular value of $\alpha$. We observe that the squeezing ($S_H<0$) depends on the angle $\phi$ and appears periodically. The degree of squeezing for a fixed angle $\phi$ (where the squeezing exists) becomes higher when $m$ is increased, i.e. when more number of photons are added to the system. In Fig. \ref{fig1}(b), we show the nature of $S_H$ in different orders $(N=2,3,4)$ for a particular value of $\phi$ and for a single PACS, where we notice that the squeezing occurs roughly for $\vert\alpha\vert > 1$. However, in this case the degree of squeezing becomes less when we increase the order $N$. Nevertheless, the most interesting observation follows from the fact that the squeezing coefficients of the NCHO (solid and dotted lines) are slightly more negative in comparison to those of the ordinary harmonic oscillator (scattered dots) in both of the figures for all cases corresponding to the same set of parameters. It means that the quadratures of the NCHO are qualitatively more squeezed and, hence, NCHO is more nonclassical than that of the harmonic oscillator.
\end{subsection}
\begin{subsection}{Hong--Mandel-type higher order squeezing}
Another type of higher order squeezing was suggested by Hong and Mandel \cite{Hong_Mandel}, who proposed that a quadrature of the form 
\begin{equation}
Y(\phi) = \frac{1}{2}\left(A_qe^{-i\phi}+A_q^\dagger e^{i\phi}\right),
\end{equation}
is said to be squeezed to the order $2N$ for the state $\vert\alpha,m\rangle_q$ for some values of $\phi$ if
\begin{equation}\label{QSqueezingHM}
{}_q\big\langle\alpha,m\big\vert (\Delta Y(\phi))^{2N}\big\vert\alpha,m\big\rangle_q < (2N-1)!!\frac{[A_q,A_q^\dagger]^N}{4^N},
\end{equation}
where $(2N-1)!!=1.3.5....(2N-1)$, and $\Delta Y(\phi)=Y(\phi)-_q\langle Y(\phi)\rangle_q$. Naturally, the degree of the Hong--Mandel-type squeezing can be computed from the squeezing coefficient $S_{HM}$
\begin{equation}\label{HM}
S_{HM} = \frac{2^{2N}{}_q\langle(\Delta Y(\phi))^{2N}\rangle_q -(2N-1)!![A_q,A_q^\dagger]^N}{(2N-1)!![A_q,A_q^\dagger]^N},
\end{equation}
where $S_{HM}<0$ implies the presence of squeezing. We would like to mention that for the case of usual harmonic oscillator satisfying $[a,a^\dagger]=1$, the computation of the squeezing coefficient \myref{HM} becomes slightly easier. However, in our case, or more precisely, for any cases where $[A,A^\dagger]\neq 1$, one needs to consider the general form of the squeezing coefficient as given in \myref{HM}. In \cite{Aeineh_Tavassoly}, the authors attempted to compute the squeezing coefficient for a nonlinear PACS by considering the commutator of the nonlinear oscillator $[A,A^\dagger]$ to be same as that of the harmonic oscillator, which we claim to be inappropriate. Nevertheless, in order to obtain a fairly general form of the squeezing coefficient $S_{HM}$ upto order $2N$, we compute
\begin{alignat}{1}
[A_q,A_q^\dagger]^N &=\left[1+(q^2-1)A_q^\dagger A_q\right]^N = \displaystyle\sum_{k=0}^{N} \begin{pmatrix} N \\ k \end{pmatrix} (q^2-1)^k(A_q^\dagger A_q)^k, \label{HM1}\\
_q\langle(\Delta Y(\phi))^{2N}\rangle_q &=\displaystyle\sum_{k=0}^{2N} \begin{pmatrix} 2N \\ k \end{pmatrix} (-1)^k{}_q\langle Y(\phi)^{2N-k}\rangle_q{}_q\langle Y(\phi)\rangle^k_q, \label{HM2}
\end{alignat}
with $A_q^\dagger A_q=[n]_q$ being the number operator of the NCHO. Using the following relations
\begin{figure}
\centering   \includegraphics[width=8.2cm,height=6.0cm]{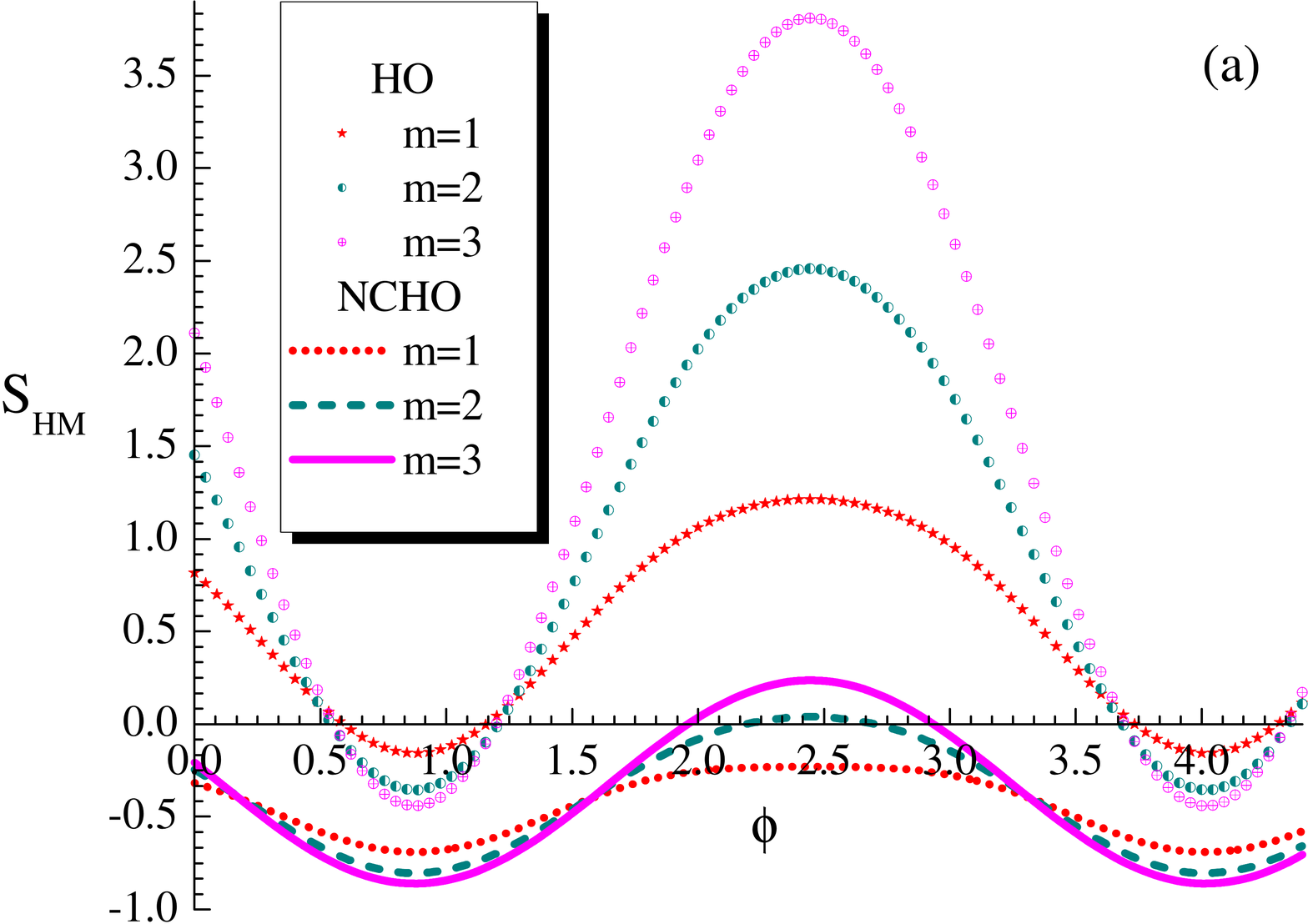}
\includegraphics[width=8.2cm,height=6.0cm]{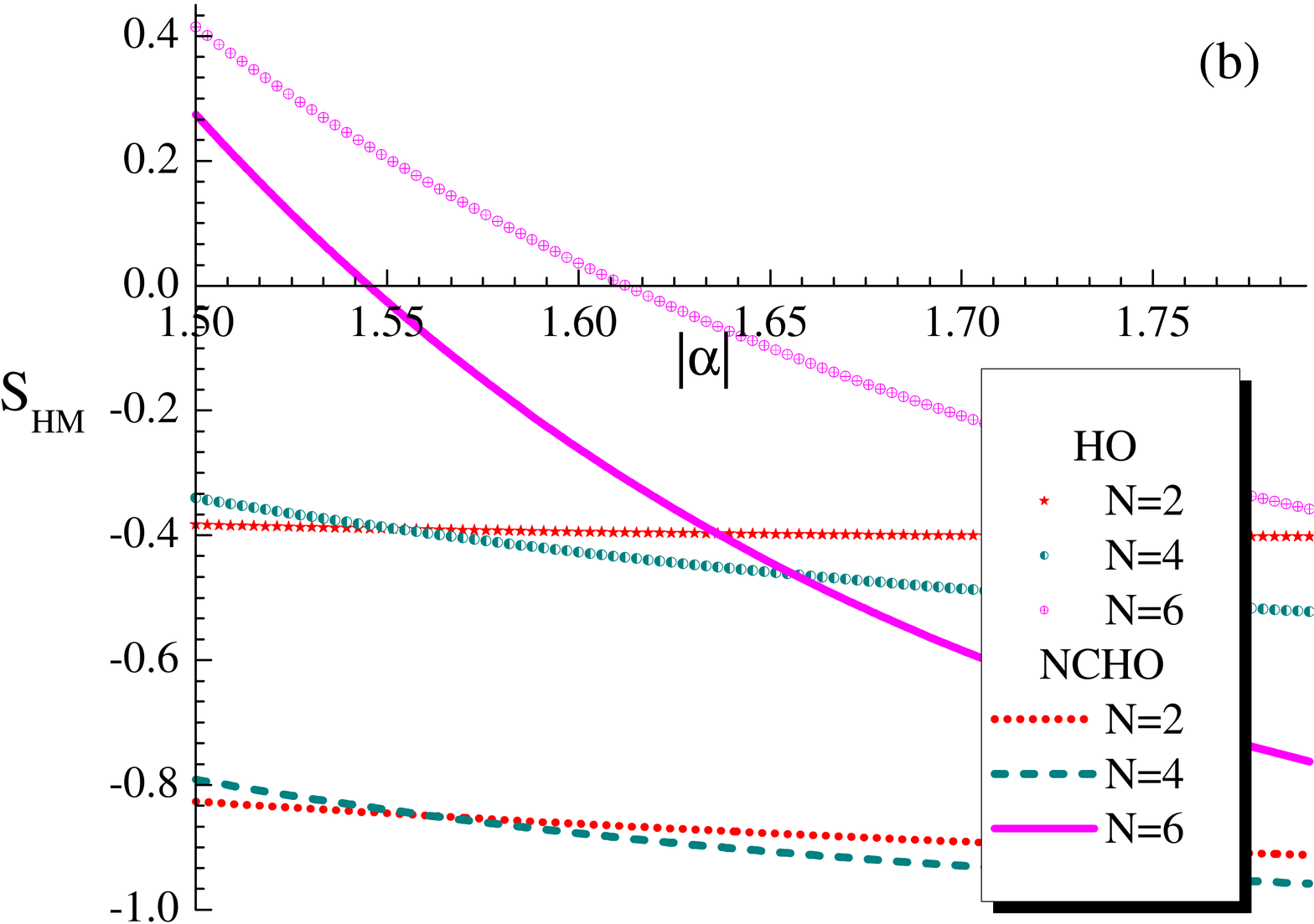}
\caption{\small{Comparison of Hong--Mandel-type higher order squeezing coefficient of a harmonic oscillator (scattered dots) versus a NCHO (for $q=0.9$) (solid and dotted lines) as a function of (a) $\phi$ for different values of $m$, with $N=4,\alpha =1.0+1.2 i$ (b) $\vert\alpha\vert$ for different values of $N$, with $\phi =0.1,m=3$.}}
\label{fig2}
\end{figure}
\begin{alignat}{1}
{}_q\langle Y(\phi)\rangle^k_q &=\displaystyle\sum_{s=0}^{k} \begin{pmatrix} k \\ s \end{pmatrix} 2^{-k}e^{i\phi(2s-k)}{}_q\langle A_q\rangle^{k-s}_q{}_q\langle A_q^\dagger\rangle^s_q, \\
{}_q\langle Y(\phi)^2\rangle_q &= \frac{1}{4}\left[1+(1+q^2){}_q\langle A_q^\dagger A_q\rangle_q+2\text{Re}\left\{{}_q\langle A_q^2\rangle_qe^{-2i\phi}\right\}\right], \\
{}_q\langle Y(\phi)^3\rangle_q &= \frac{1}{4}\text{Re}\left[{}_q\langle A_q^3\rangle_qe^{-3i\phi}+(2+q^2){}_q\langle A_q\rangle_qe^{-i\phi}+(1+q^2+q^4){}_q\langle A_q^\dagger A_q^2\rangle_qe^{-i\phi}\right], \\
{}_q\langle Y(\phi)^4\rangle_q &= \frac{1}{16}\left[2+q^2+(3+5q^2+3q^4+q^6){}_q\langle A_q^\dagger A_q\rangle_q+(\mu+q^4){}_q\langle A_q^{\dagger 2} A_q^2\rangle_q\right. \\
&~~~\left.+2\text{Re}\left\{{}_q\langle A_q^4\rangle_qe^{-4i\phi}+(3+2q^2+q^4){}_q\langle A_q^2\rangle_qe^{-2i\phi}+(\mu-q^8){}_q\langle A_q^\dagger A_q^3\rangle_qe^{-2i\phi}\right\}\right], \notag \\
{}_q\langle Y(\phi)^5\rangle_q &= \frac{1}{16}\text{Re}\left[{}_q\langle A_q^5\rangle_qe^{-5i\phi}+\left\{\mu{}_q\langle A_q^\dagger A_q^4\rangle_q+(4+3q^2+2q^4+q^6){}_q\langle A_q^3\rangle_q\right\}e^{-3i\phi}\right. \notag \\
&~~~\left.+\left\{(5+6q^2+3q^4+q^6){}_q\langle A_q\rangle_q+(1+3\mu+4q^2+6q^4+3q^6+q^{10}){}_q\langle A_q^\dagger A_q^2\rangle_q\right.\right. \notag \\
&~~~\left.\left.+(\mu+q^4+q^6+q^8+q^{10}+q^{12}){}_q\langle A_q^{\dagger 2} A_q^3\rangle_q\right\}e^{-i\phi}\right], \\
{}_q\langle Y(\phi)^6\rangle_q &= \frac{1}{64}\left[5+6q^2+3q^4+q^6+(9+22q^2+25q^4+19q^6+10q^8+4q^{10}+q^{12}){}_q\langle A_q^\dagger A_q\rangle_q\right. \notag \\
&~~~+\left.(5+9q^2+17q^4+18(q^6+q^8)+12q^{10}+7q^{12}+3q^{14}+q^{16}){}_q\langle A_q^{\dagger 2} A_q^2\rangle_q+(\lambda+q^6+q^{10} \right. \notag \\
&~~~+\left.q^{12}+q^{14}+q^{18}){}_q\langle A_q^{\dagger 3} A_q^3\rangle_q+2\text{Re}\left\{{}_q\langle A_q^6\rangle_qe^{-6i\phi}+\left((5+4q^2+3q^4+2q^6+q^8){}_q\langle A_q^4\rangle_q\right.\right.\right. \notag \\
&~~~+\left.\left.\left.(\mu+q^{10}){}_q\langle A_q^{\dagger} A_q^5\rangle_q\right)e^{-4i\phi}+\left((9+13q^2+12q^4+7q^6+3q^8+q^{10}){}_q\langle A_q^2\rangle_q+(5+9q^2 \right.\right.\right. \notag \\
&~~~+\left.\left.\left.12q^4+14q^6+10q^8+6q^{10}+3q^{12}+q^{14}){}_q\langle A_q^\dagger A_q^3\rangle_q+\lambda{}_q\langle A_q^{\dagger 2}A_q^4\rangle_q\right)e^{-2i\phi}\right\}\right], 
\end{alignat}
and Eq. \myref{ADNAL}, \myref{AAD} one can easily evaluate \myref{HM2}, which when replaced in \myref{HM} one obtains a complete expression of the Hong--Mandel squeezing coefficient $S_{HM}$ explicitly upto order $6$, where $\lambda=\mu+q^4+q^6+2q^8+2q^{10}+2q^{12}+q^{14}+q^{16}$ and $\mu=1+q^2+q^4+q^6+q^8$. In Fig. \ref{fig2}, the nature of the Hong--Mandel-type squeezing coefficient \myref{HM} is shown as functions of $\phi$ and $\alpha$ in panels (a) and (b), respectively. The qualitative behaviour of the plots in Fig. \ref{fig2} remains similar to that of Fig. \ref{fig1}, such as, the degree of squeezing is enhanced when we add more photons for a fixed angle $\phi$ in panel (a) and it is reduced when we increase the order $N$ for a fixed value of $\alpha$ in panel (b). Also, the squeezing in quadrature of the NCHO stays higher than that of the harmonic oscillator.
\end{subsection}
\end{section}
\begin{section}{Higher order sub-Poissonian photon statistics}\label{sec4}
Let us now study the higher order photon statistics of the deformed PACS \myref{qPACS} and check whether it is sub-Poissonian or not. The concept of higher order sub-Poissonian photon statistics was introduced in \cite{Lee1,Kim,Erenso_Vyas_Singh} in terms of factorial moment, and was improved later by many authors; such as, \cite{Kim_Yoon,Prakash_Mishra,Duc_Noh}. Some of them studied the higher order Mandel parameter, while some other used the higher order correlation function for the purpose of testing the nature of the photon statistics. However, the method discussed in \cite{Lee1,Kim,Erenso_Vyas_Singh,Kim_Yoon,Prakash_Mishra,Duc_Noh} does not work in our case, since we are not working on the usual quantum mechanical systems satisfying $[a,a^\dagger]=1$, but a different system obeying the $q$-deformed algebra \myref{qDeformed}. Indeed, in \cite{Aeineh_Tavassoly} the authors used those standard relations and attempted inappropriately to study the higher order sub-Poissonian statistics for nonlinear systems. Instead, here we start with the original definition of the second order correlation function (for zero delay time) $g^{(2)}(0)$ introduced in \cite{Glauber2}, and generalise it to an arbitrary order $N$ as follows
\begin{equation}\label{qCORF}
g^{(N)}(0) = \frac{_q\langle(\Delta M)^N\rangle_q-{}_q\langle M\rangle_q}{~_q\langle M\rangle^N_q}+1.
\end{equation} 
We also generalise the introductory definition of the Mandel parameter $Q$ \cite{Mandel} to an arbitrary order $N$ as given by
\begin{figure}[h]
\centering   \includegraphics[width=8.2cm,height=6.0cm]{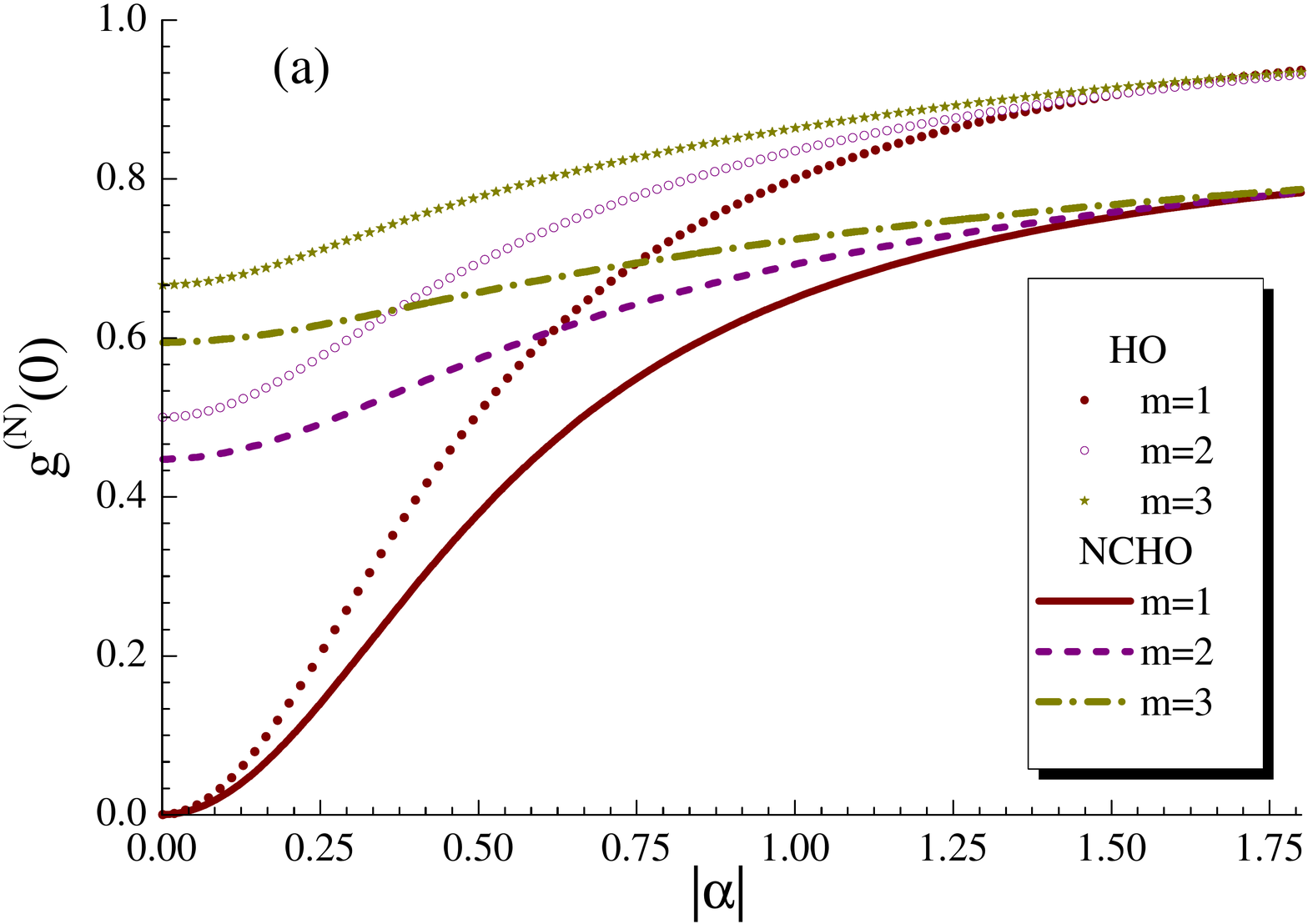}
\includegraphics[width=8.2cm,height=6.0cm]{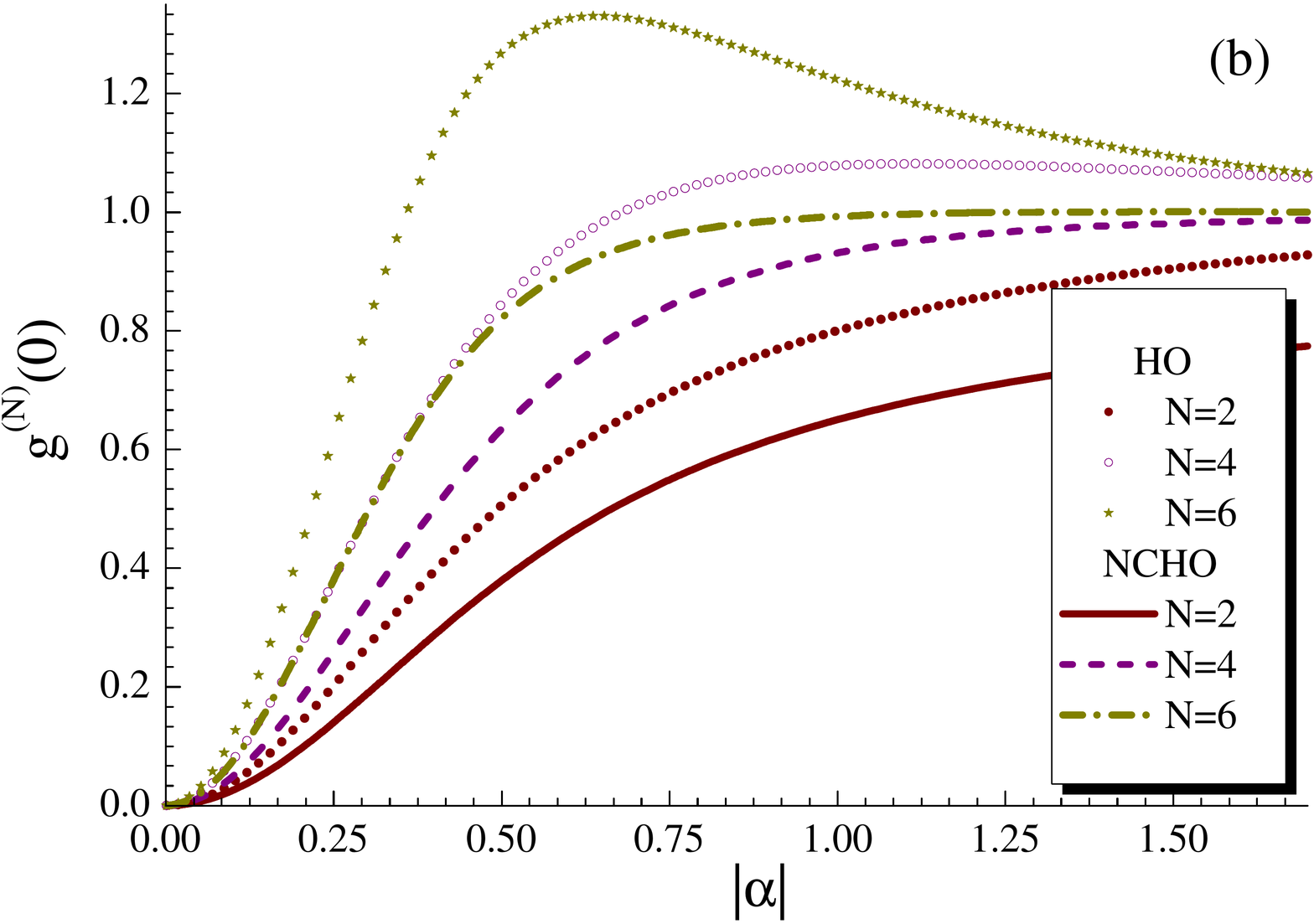}
\caption{\small{Comparison of higher order correlation function of a harmonic oscillator (scattered dots) versus a NCHO (for $q=0.9$) (solid and dotted lines) as a function of $\vert\alpha\vert$ (a) for different values of $m$, with $N=2$ (b) for different values of $N$, with $m=1$.}}
\label{fig3}
\end{figure}
\begin{figure}
\centering   \includegraphics[width=8.2cm,height=6.0cm]{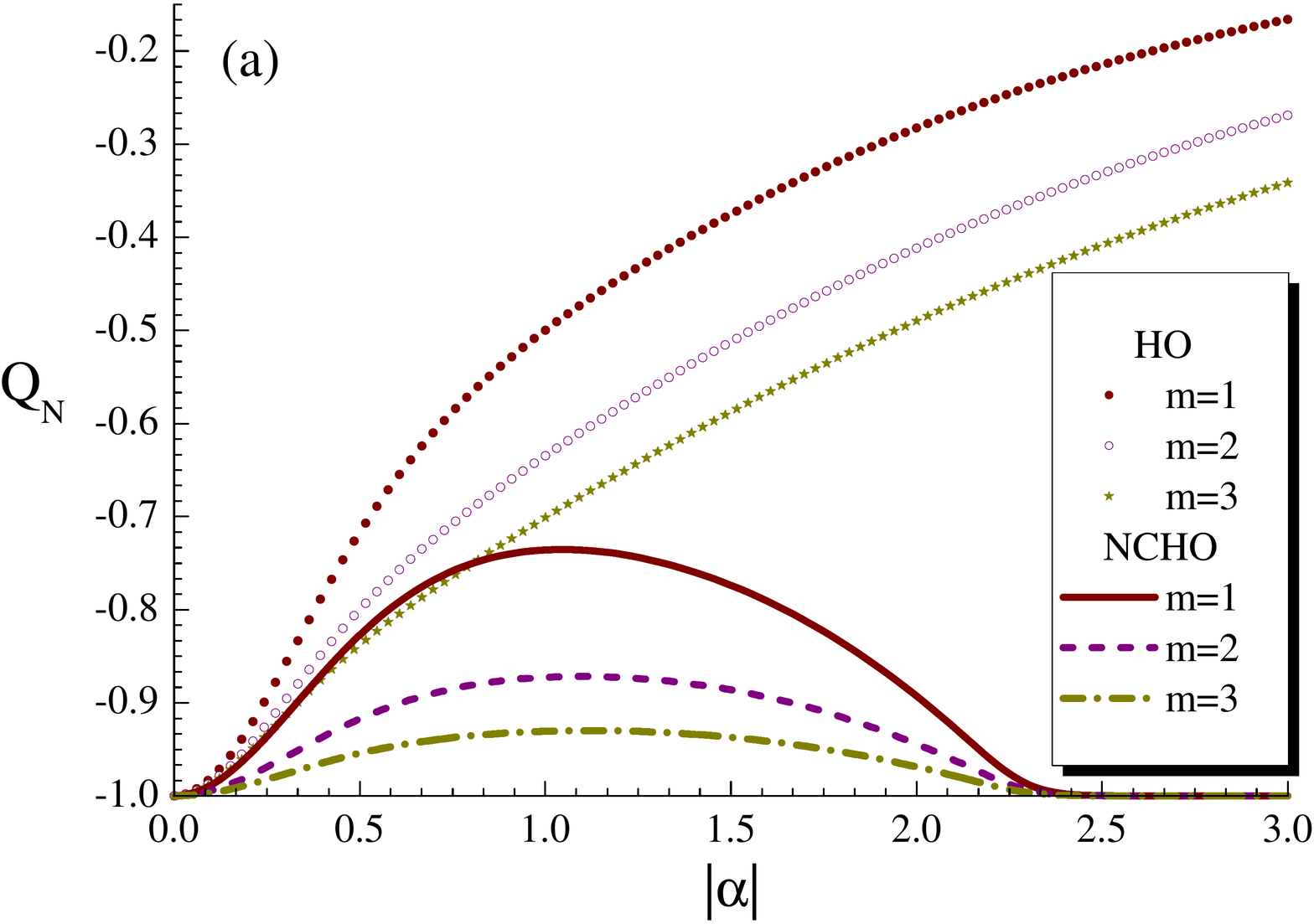}
\includegraphics[width=8.2cm,height=6.0cm]{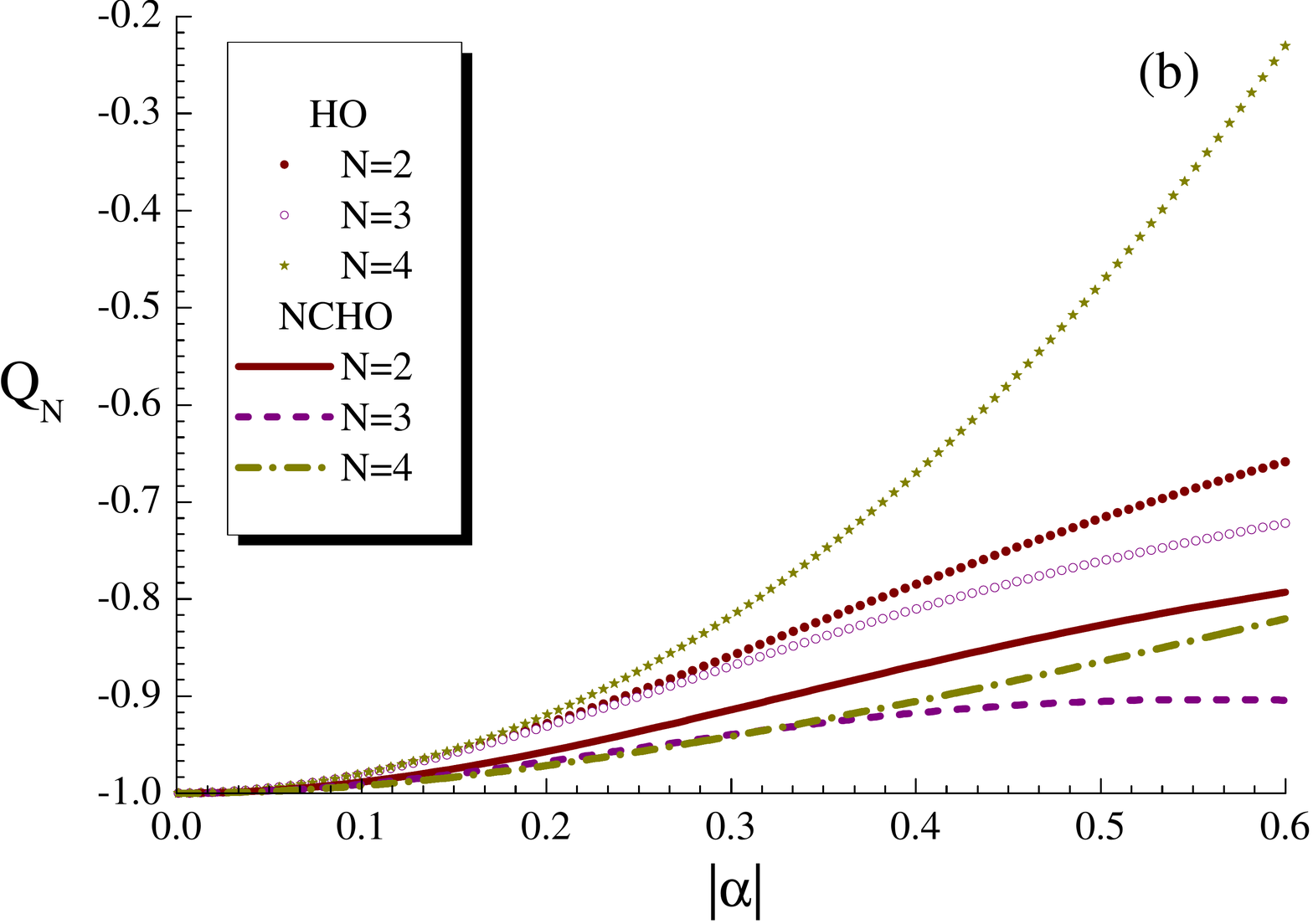}
\caption{\small{Comparison of higher order Mandel parameter of a harmonic oscillator (scattered dots) versus a NCHO (for $q=0.9$) (solid and dotted lines) as a function of $\vert\alpha\vert$ (a) for different values of $m$, with $N=2$ (b) for different values of $N$, with $m=1$.}}
\label{fig4}
\end{figure}
\begin{equation}\label{qMandel}
Q_N = \frac{_q\langle(\Delta M)^N\rangle_q}{~_q\langle M\rangle_q}-1,
\end{equation}
where $\Delta M=A_q^\dagger A_q-{}_q\langle A_q^\dagger A_q\rangle_q$ is the dispersion of the number operator $M=A_q^\dagger A_q$. It is well-known that the definitions \myref{qCORF} and \myref{qMandel} are equivalent while testing the nature of photon statistics of the system. Or in other words, for all $N\geq 2$ the photon number distribution is Poissonian if $Q_N=0$ (and $g^{(N)}(0)=1$). Whereas, $Q_N>0~(g^{(N)}(0)>1)$ and $Q_N<0~(g^{(N)}(0)<1)$ correspond to the super-Poissonian (photon bunching) and sub-Poissonian (photon anti-bunching) cases, respectively. Although it seems that either the correlation function or the Mandel parameter would be sufficient to study for the purpose of testing the squeezing behaviour of photon number, however, it is clear from \myref{qCORF} and \myref{qMandel} that they are not trivially connected to each other for arbitrary orders. That is why we analyse both of them. Using the following identity
\begin{equation}
_q\langle(\Delta M)^N\rangle_q = \displaystyle\sum_{k=0}^{N} \begin{pmatrix} N \\ k \end{pmatrix} (-1)^k{}_q\langle (A^\dagger_qA_q)^{N-k}\rangle_q{}_q\langle A^\dagger_qA_q\rangle^k_q,
\end{equation}
with\\
\begin{equation}\label{Eq44}
_q\langle(A_q^\dagger A_q)^N\rangle_q = \frac{1}{\mathcal{\hat{N}}^2}\displaystyle\sum_{n=0}^{\infty}\frac{\vert\alpha\vert^{2n}[n+m]_q!}{([n]_q!)^2}[n+m]_q^N,
\end{equation}
and the expectation value of the number operator $_q\langle A_q^\dagger A_q\rangle_q$, which is obtained by choosing $N=1$ in \myref{Eq44}, we can evaluate the higher order correlation function $g^{(N)}(0)$ \myref{qCORF} and the higher order Mandel parameter $Q_N$ \myref{qMandel} explicitly upto order $N$. Fig. \ref{fig3}(a) shows the behaviour of the correlation function \myref{qCORF} with respect to $\vert\alpha\vert$ for different values of added photon numbers $m$. It is clear that with the increase of $m$ the value of $g^{(N)}(0)$ increases, i.e. the photon number squeezing becomes less pronounced when $m$ is increased. Fig. \ref{fig3}(b) demonstrates the dependence of $g^{(N)}(0)$ for different orders of squeezing $(N=2,4,6)$ and a similar effect happens in this case as well, i.e. $g^{(N)}(0)$ increases with the increase of the order $N$, and for bigger values of $\vert\alpha\vert$ it eventually becomes higher than $1$, for $N>4$ roughly. Thus, the lower orders perform better and provide better squeezing than the higher orders. However, we can still see the squeezing effect for higher orders when $\vert\alpha\vert$ remains small. The sub-Poissonian character with respect to the higher order Mandel parameter $Q_N$ is shown in Fig. \ref{fig4}, which shows that $Q_N$ stays negative when we choose the parameters appropriately for different orders $N$ and for different photon numbers $m$. Nevertheless, when we study the NCHO (solid and dotted lines), the whole analysis of higher order sub-Poissonian behaviour for correlation function $g^{(N)}(0)$ and Mandel parameter $Q_N$ becomes much improved. Moreover, the entire behaviour can be enhanced by controlling the noncommutative parameter $q$. As for instance as Fig. \ref{fig5} shows that the Mandel parameter becomes more and more negative when we go beyond the harmonic oscillator limit by increasing the noncommutativity, $\vert q\vert <1$. However, this is just an example. Actually, it occurs for every case that we have discussed in our article, which we do not present here. 
\begin{figure}
\centering   \includegraphics[width=8.5cm,height=6.0cm]{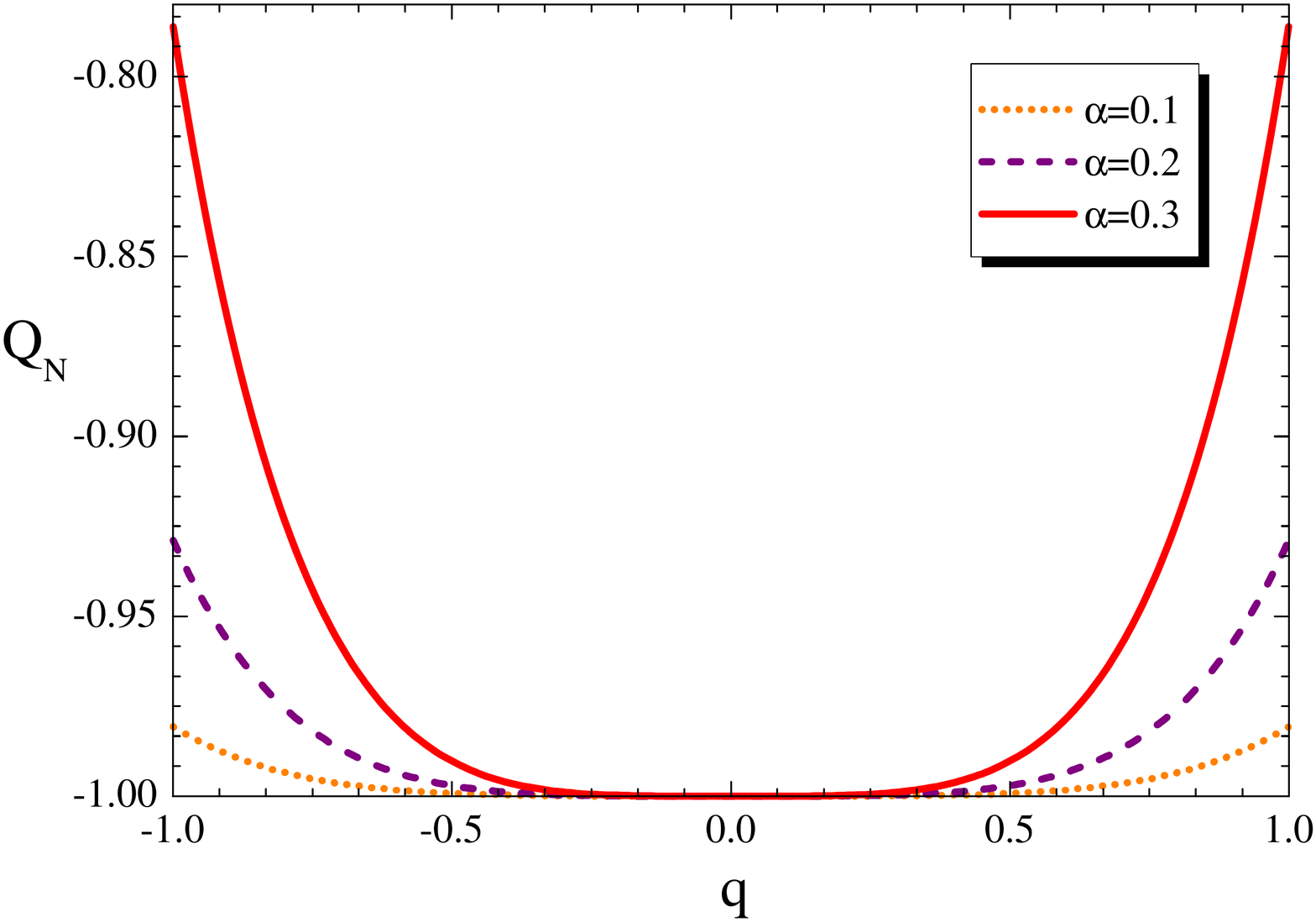}
\caption{\small{Higher order Mandel parameter of a NCHO as a function of $q$ for different values of $\alpha$, with $m=1$ and $N=2$.}}
\label{fig5}
\end{figure}
\end{section}
\begin{section}{Conclusions}\label{sec5}
In summary, we have presented a formalism for the construction of PACS for a NCHO associated to a $q$-deformed oscillator algebra \myref{qPACS}, which in a special limit ($q\rightarrow 1$) reduce to the PACS of the harmonic oscillator. Nonclassical properties of the corresponding system have been analysed in detail by studying several types of quadrature (Hillery type and Hong--Mandel-type) and photon number (Mandel parameter and correlation function) squeezing method. In particular, we provide general expressions of the quadrature squeezing coefficients (Hillery and Hong--Mandel), Mandel parameter and correlation function in arbitrary orders $N$, which can be utilised to study the nonclassical properties of similar type of systems. Qualitative comparisons of our results emerging out of the NCHO with that of the usual harmonic oscillator have been reported alongside. Throughout the analysis we observe an improved degree of squeezing for noncommutative case in comparison to usual system for the choice of same set of parameters. Moreover, the entire behaviour of squeezing can be enriched by increasing the noncommutativity of the underlying system by controlling the noncommutative parameter $q$. Since, there exists a Hermitian representation of the corresponding system, it raises a natural question whether such type of systems are possible to implement in quantum optics. If so, one might obtain an extra degree of freedom for controlling the squeezing behaviour of the PACS. However, the whole analysis remains a theoretical prediction only and, thus, it would be interesting to explore such type of models further in the context of quantum optics, quantum information and computation.  

\vspace{0.5cm} \noindent \textbf{\large{Acknowledgements:}} SD is supported by the Postdoctoral Fellowship jointly funded by the Laboratory of Mathematical Physics of the Centre de Recherches Math{\'e}matiques and by Prof. Syed Twareque Ali, Prof. Marco Bertola and Prof. V{\'e}ronique Hussin.

\end{section}


\begin{thebibliography}{100}

\bibitem{Glauber2}
R.~J. Glauber,
\newblock The quantum theory of optical coherence,
\newblock Phys. Rev. \textbf{130}, 2529 (1963).

\bibitem{Sudarshan2}
E.~C.~G. Sudarshan,
\newblock Equivalence of semiclassical and quantum mechanical descriptions of statistical light beams,
\newblock Phys. Rev. Lett. \textbf{10}, 277--279 (1963).

\bibitem{Agarwal_Tara}
G.~S. Agarwal and K.~Tara,
\newblock Nonclassical properties of states generated by the excitations on a coherent state,
\newblock Phys. Rev. A \textbf{43}, 492 (1991).

\bibitem{Dodonov_etal}
V.~V. Dodonov, M.~A. Marchiolli, Y.~A. Korennoy, V.~I. Man’ko and Y.~A. Moukhin,
\newblock Dynamical squeezing of photon-added coherent states,
\newblock Phys. Rev. A \textbf{58}, 4087 (1998).

\bibitem{Sivakumar_PACS}
S.~Sivakumar,
\newblock Photon-added coherent states as nonlinear coherent states,
\newblock J. Phys. A \textbf{32}, 3441 (1999).

\bibitem{Sixdeniers_Penson}
J.~M. Sixdeniers and K.~A. Penson,
\newblock On the completeness of photon-added coherent states,
\newblock J. Phys. A \textbf{34}, 2859 (2001).

\bibitem{Daoud}
M.~Daoud,
\newblock Photon-added coherent states for exactly solvable {Hamiltonians},
\newblock Phys. Lett. A \textbf{305}, 135--143 (2002).

\bibitem{Popov}
D.~Popov,
\newblock Photon-added {Barut--Girardello coherent states of the pseudoharmonic oscillator},
\newblock J. Phys. A \textbf{35}, 7205 (2002).

\bibitem{Naderi_etal}
M.~H. Naderi, M.~Soltanolkotabi and R.~Roknizadeh,
\newblock New photon-added and photon-depleted coherent states associated with inverse $q$-boson operators: nonclassical properties,
\newblock J. Phys. A \textbf{37}, 3225 (2004).

\bibitem{Sudheesh_etal}
C.~Sudheesh, S.~Lakshmibala and V.~Balakrishnan,
\newblock Wave packet dynamics of photon-added coherent states,
\newblock Europhys. Lett. \textbf{71}, 744 (2005).

\bibitem{Sudheesh_etal1}
C.~Sudheesh, S.~Lakshmibala and V.~Balakrishnan,
\newblock Squeezing and higher-order squeezing of photon-added coherent states propagating in a {Kerr-like medium},
\newblock J. Opt. B \textbf{7}, S728 (2005).

\bibitem{Duc_Noh}
T.~M. Duc and J.~Noh,
\newblock Higher-order properties of photon-added coherent states,
\newblock Opt. Commun. \textbf{281}, 2842--2848 (2008).

\bibitem{Gorska_Penson_Duchamp}
K.~Gorska, K.~A. Penson and G.~H.~E. Duchamp,
\newblock Generation of coherent states of photon-added type via pathway of eigenfunctions,
\newblock J. Phys. A \textbf{43}, 375303 (2010).

\bibitem{Safaeian_Tavassoly}
O.~Safaeian and M.~K. Tavassoly,
\newblock Deformed photon-added nonlinear coherent states and their non-classical properties,
\newblock J. Phys. A \textbf{44}, 225301 (2011).

\bibitem{Mojaveri_Dehghani_Mahmoodi}
B.~Mojaveri, A.~Dehghani and S.~Mahmoodi,
\newblock New class of generalized photon-added coherent states and some of their non-classical properties,
\newblock Phys. Scr. \textbf{89}, 085202 (2014).

\bibitem{Mojaveri_Dehghani_Mohammadzadeh}
B.~Mojaveri, A.~Dehghani and B.~Ali-Mohammadzadeh,
\newblock {Even and Odd Deformed Photon Added Nonlinear Coherent States},
\newblock Int. J. Theor. Phys. \textbf{55}, 421--431 (2016).

\bibitem{Li_Jing_Zhan}
Y.~Li, H.~Jing and M.-S. Zhan,
\newblock Optical generation of a hybrid entangled state via an entangling single-photon-added coherent state,
\newblock J. Phys. B \textbf{39}, 2107 (2006).

\bibitem{Berrada_etal}
K.~Berrada, S.~Abdel-Khalek, H.~Eleuch and Y.~Hassouni,
\newblock Beam splitting and entanglement generation: excited coherent states,
\newblock Quant. Inf. Process. \textbf{12}, 69--82 (2013).

\bibitem{Stoler}
D.~Stoler,
\newblock Equivalence classes of minimum uncertainty packets,
\newblock Phys. Rev. D \textbf{1}, 3217 (1970).

\bibitem{Hollenhorst}
J.~N. Hollenhorst,
\newblock Quantum limits on resonant-mass gravitational-radiation detectors,
\newblock Phys. Rev. D \textbf{19}, 1669 (1979).

\bibitem{Walls}
D.~F. Walls,
\newblock Squeezed states of light,
\newblock Nature (London) \textbf{306}, 141 (1983).

\bibitem{Dodonov_Malkin_Manko}
V.~V. Dodonov, I.~A. Malkin and V.~I. Man'Ko,
\newblock Even and odd coherent states and excitations of a singular oscillator,
\newblock Physica \textbf{72}, 597--615 (1974).

\bibitem{Xia_Guo}
Y.~Xia and G.~Guo,
\newblock Nonclassical properties of even and odd coherent states,
\newblock Phys. Lett. A \textbf{136}, 281--283 (1989).

\bibitem{Agarwal_Biswas}
G.~S. Agarwal and A.~Biswas,
\newblock Quantitative measures of entanglement in pair-coherent states,
\newblock J. Opt. B \textbf{7}, 350--354 (2005).

\bibitem{Stoler_Saleh_Teich}
D.~Stoler, B.~E.~A. Saleh and M.~C. Teich,
\newblock Binomial states of the quantized radiation field,
\newblock J. Mod. Opt. \textbf{32}, 345--355 (1985).

\bibitem{Lee}
C.~T. Lee,
\newblock Photon antibunching in a free-electron laser,
\newblock Phys. Rev. A \textbf{31}, 1213 (1985).

\bibitem{Zavatta_Viciani_Bellini}
A.~Zavatta, S.~Viciani and M.~Bellini,
\newblock Quantum-to-classical transition with single-photon-added coherent states of light,
\newblock Science \textbf{306}, 660--662 (2004).

\bibitem{Barbieri_etal}
M.~Barbieri, N.~Spagnolo, M.~G. Genoni, F.~Ferreyrol, R.~Blandino, M.~G.~A. Paris, P.~Grangier and R.~Tualle-Brouri,
\newblock {Non-Gaussianity of quantum states: An experimental test on single-photon-added coherent states},
\newblock Phys. Rev. A \textbf{82}, 063833 (2010).

\bibitem{Kalamidas_Gerry_Benmoussa}
D.~Kalamidas, C.~C. Gerry and A.~Benmoussa,
\newblock Proposal for generating a two-photon added coherent state via down-conversion with a single crystal,
\newblock Phys. Lett. A \textbf{372}, 1937--1940 (2008).

\bibitem{Arik_Coon}
M.~Arik and D.~D. Coon,
\newblock Hilbert spaces of analytic functions and generalized coherent states,
\newblock J. Math. Phys. \textbf{17}, 524--527 (1976).

\bibitem{Biedenharn}
L.~C. Biedenharn,
\newblock The quantum group {$SUq (2)$} and a $q$-analogue of the boson operators,
\newblock J. Phys. A \textbf{22}, L873 (1989).

\bibitem{Macfarlane}
A.~J. Macfarlane,
\newblock On $q$-analogues of the quantum harmonic oscillator and the quantum group {$SU(2)q$},
\newblock J. Phys. A \textbf{22}, 4581 (1989).

\bibitem{Sun_Fu}
C.-P. Sun and H.-C. Fu,
\newblock The $q$-deformed boson realisation of the quantum group {$SU(n)q$} and its representations,
\newblock J. Phys. A \textbf{22}, L983 (1989).

\bibitem{Kulish_Damaskinsky}
P.~P. Kulish and E.~V. Damaskinsky,
\newblock On the $q$ oscillator and the quantum algebra {$SUq(1, 1)$},
\newblock J. Phys. A \textbf{23}, L415 (1990).

\bibitem{Bagchi_Fring}
B.~Bagchi and A.~Fring,
\newblock Minimal length in quantum mechanics and {non-Hermitian Hamiltonian systems},
\newblock Phys. Lett. A \textbf{373}, 4307--4310 (2009).

\bibitem{Dey_Fring_Gouba}
S.~Dey, A.~Fring and L.~Gouba,
\newblock $\mathcal{PT}$-symmetric non-commutative spaces with minimal volume uncertainty relations,
\newblock J. Phys. A \textbf{45}, 385302 (2012).

\bibitem{Kempf_Mangano_Mann}
A.~Kempf, G.~Mangano and R.~B. Mann,
\newblock Hilbert space representation of the minimal length uncertainty relation,
\newblock Phys. Rev. D \textbf{52}, 1108 (1995).

\bibitem{Dey_Fring_Gouba_Castro}
S.~Dey, A.~Fring, L.~Gouba and P.~G. Castro,
\newblock Time-dependent $q$-deformed coherent states for generalized uncertainty relations,
\newblock Phys. Rev. D \textbf{87}, 084033 (2013).

\bibitem{Bender_Boettcher}
C.~Bender and S. Boettcher,
\newblock Real spectra in {non-Hermitian Hamiltonians having $\mathcal{PT}$-symmetry},
\newblock Phys. Rev. Lett. \textbf{80}, 5243 (1998).

\bibitem{Mostafazadeh_2002}
A.~Mostafazadeh,
\newblock {Pseudo-Hermiticity versus $\mathcal{PT}$-symmetry: the necessary condition for the reality of the spectrum of a non-Hermitian Hamiltonian},
\newblock J. Math. Phys. \textbf{43}, 205--214 (2002).

\bibitem{Dey_Fring_Khantoul}
S.~Dey, A.~Fring and B. Khantoul,
\newblock Hermitian versus {non-Hermitian} representations for minimal length uncertainty relations,
\newblock J. Phys. A \textbf{46}, 335304 (2013).

\bibitem{El-Ganainy}
R.~El-Ganainy, K. G.~Makris, D. N.~Christodoulides and Z. H.~Musslimani,
\newblock Theory of coupled optical $\mathcal{PT}$-symmetric structures,
\newblock Opt. Lett. \textbf{32}, 2632--2634 (2007).

\bibitem{Guo}
A.~Guo, G. J.~Salamo, D.~Duchesne, R.~Morandotti, M. Volatier-Ravat, V. Aimez, G. A. Siviloglou and D. N.~Christodoulides,
\newblock Observation of $\mathcal{PT}$-symmetry breaking in complex optical potentials,
\newblock Phys. Rev. Lett. \textbf{103}, 093902 (2009).



\bibitem{Longhi}
S.~Longhi
\newblock $\mathcal{PT}$-symmetric laser absorber,
\newblock Phys. Rev. A \textbf{82}, 031801 (2010).

\bibitem{Chong_Ge_Cao_Stone}
Y. D.~Chong, L.~Ge, H.~Cao and A. D. Stone,
\newblock Coherent perfect absorbers: time-reversed lasers,
\newblock Phys. Rev. Lett. \textbf{105}, 053901 (2010).

\bibitem{Manko_Marmo_Sudarshan_Zaccaria}
V.~I. Man'ko, G.~Marmo, E.~C.~G. Sudarshan and F.~Zaccaria,
\newblock $f$-oscillators and nonlinear coherent states,
\newblock Phys. Scr. \textbf{55}, 528 (1997).

\bibitem{Quesne}
C.~Quesne,
\newblock New $q$-deformed coherent states with an explicitly known resolution of unity,
\newblock J. Phys. A \textbf{35}, 9213 (2002).

\bibitem{Dey_Fring_squeezed}
S.~Dey and A.~Fring,
\newblock Squeezed coherent states for noncommutative spaces with minimal length uncertainty relations,
\newblock Phys. Rev. D \textbf{86}, 064038 (2012).

\bibitem{Mancini}
S.~Mancini,
\newblock Even and odd nonlinear coherent states,
\newblock Phys. Lett. A \textbf{233}, 291--296 (1997).

\bibitem{Osland}
P.~Osland and J.-z. Zhang,
\newblock Critical phenomenon of a consistent $q$-deformed squeezed state,
\newblock Ann. Phys. \textbf{290}, 45--52 (2001).

\bibitem{Dey}
S.~Dey,
\newblock $q$-deformed noncommutative cat states and their nonclassical properties,
\newblock Phys. Rev. D \textbf{91}, 044024 (2015).

\bibitem{Dey_Fring_Hussin}
S.~Dey, A.~Fring and V.~Hussin,
\newblock Nonclassicality versus entanglement in a noncommutative space,
\newblock arXiv:1506.08901.

\bibitem{Fakhri_Hashemi}
H.~Fakhri and A.~Hashemi,
\newblock {Nonclassical properties of the $q$-coherent and $q$-cat states of the Biedenharn-Macfarlane $q$ oscillator with $q> 1$},
\newblock Phys. Rev. A \textbf{93}, 013802 (2016).

\bibitem{Dey_Hussin}
S.~Dey and V.~Hussin,
\newblock Entangled squeezed states in noncommutative spaces with minimal length uncertainty relations,
\newblock Phys. Rev. D \textbf{91}, 124017 (2015).

\bibitem{Hillery}
M.~Hillery,
\newblock Amplitude-squared squeezing of the electromagnetic field,
\newblock Phys. Rev. A \textbf{36}, 3796 (1987).

\bibitem{Hong_Mandel}
C.~K. Hong and L.~Mandel,
\newblock Higher-order squeezing of a quantum field,
\newblock Phys. Rev. Lett. \textbf{54}, 323 (1985).

\bibitem{Gerry_Rodrigues}
C.~Gerry and S.~Rodrigues,
\newblock Higher-order squeezing from an anharmonic oscillator,
\newblock Phys. Rev. A \textbf{35}, 4440 (1987).

\bibitem{Zhang_Xu_Chai_Li}
Z.-M. Zhang, L.~Xu, J.-L. Chai and F.-L. Li,
\newblock A new kind of higher-order squeezing of radiation field,
\newblock Phys. Lett. A \textbf{150}, 27--30 (1990).

\bibitem{Zhan}
Y.-B. Zhan,
\newblock Amplitude-cubed squeezing in harmonic generations,
\newblock Phys. Lett. A \textbf{160}, 498--502 (1991).

\bibitem{Du_Gong}
S.-D. Du and C.-D. Gong,
\newblock Squeezing of the $k$th power of the field amplitude,
\newblock Phys. Lett. A \textbf{168}, 296--300 (1992).

\bibitem{An}
N.~B. An,
\newblock Higher-order amplitude squeezing of photons propagating through a semiconductor,
\newblock Phys. Lett. A \textbf{234}, 45--52 (1997).

\bibitem{Prakash_Mishra}
H.~Prakash and D.~K. Mishra,
\newblock Higher order sub-poissonian photon statistics and their use in detection of {Hong and Mandel} squeezing and amplitude-squared squeezing,
\newblock J. Phys. B \textbf{39}, 2291 (2006).

\bibitem{Verma_Pathak}
A.~Verma and A.~Pathak,
\newblock Generalized structure of higher order nonclassicality,
\newblock Phys. Lett. A \textbf{374}, 1009--1020 (2010).

\bibitem{Aeineh_Tavassoly}
N.~Aeineh and M.~K. Tavassoly,
\newblock {Higher-Orders of Squeezing, Sub-Poissonian Statistics and Anti-Bunching of Deformed Photon-Added Coherent States},
\newblock Rep. Math. Phys. \textbf{76}, 75--89 (2015).

\bibitem{Lee1}
C.~T. Lee,
\newblock Higher-order criteria for nonclassical effects in photon statistics,
\newblock Phys. Rev. A \textbf{41}, 1721 (1990).

\bibitem{Kim}
K.~Kim,
\newblock Higher order sub-poissonian,
\newblock Phys. Lett. A \textbf{245}, 40--42 (1998).

\bibitem{Erenso_Vyas_Singh}
D.~Erenso, R.~Vyas and S.~Singh,
\newblock Higher-order sub-poissonian photon statistics in terms of factorial moments,
\newblock J. Opt. Soc. Am. B \textbf{19}, 1471--1475 (2002).

\bibitem{Kim_Yoon}
Y.~Kim and T.~H. Yoon,
\newblock Higher order sub-poissonian photon statistics of light,
\newblock Opt. Commun. \textbf{212}, 107--114 (2002).

\bibitem{Mandel}
L.~Mandel,
\newblock {Sub-Poissonian} photon statistics in resonance fluorescence,
\newblock Opt. Lett. \textbf{4}, 205--207 (1979).

\end{thebibliography}

\end{document}